\documentclass[conference]{IEEEtran}
\IEEEoverridecommandlockouts
\usepackage{cite}
\usepackage{amsmath,amssymb,amsfonts}
\usepackage{algorithmic}
\usepackage{graphicx}
\usepackage{textcomp}
\usepackage{xspace}
\usepackage{algorithm} 
\usepackage{caption}
\usepackage{threeparttable}
\usepackage[dvipsnames]{xcolor}
\usepackage{multirow}
\usepackage{hyperref}
\usepackage{tikz}
\newcommand*\emptycirc[1][0.8ex]{\tikz\draw[thick] (0,0) circle (#1);} 
\newcommand*\halfcirc[1][0.8ex]{%
  \begin{tikzpicture}
  \draw[fill] (0,0)-- (90:#1) arc (90:270:#1) -- cycle ;
  \draw[thick] (0,0) circle (#1);
  \end{tikzpicture}}
\newcommand*\fullcirc[1][0.8ex]{\tikz\fill (0,0) circle (#1);} 

\newcommand*\circled[1]{\tikz[baseline=(char.base)]{
            \node[shape=circle,draw,inner sep=0.1pt] (char) {#1};}}

\newcommand*\smallcircled[1]{\tikz[baseline=(char.base)]{
            \node[shape=circle,draw,inner sep=-0.4pt] (char) {#1};}}

\def\BibTeX{{\rm B\kern-.05em{\sc i\kern-.025em b}\kern-.08em
    T\kern-.1667em\lower.7ex\hbox{E}\kern-.125emX}}

\makeatletter
\newcommand{\linebreakand}{%
  \end{@IEEEauthorhalign}
  \hfill\mbox{}\par
  \mbox{}\hfill\begin{@IEEEauthorhalign}
}
\makeatother

\begin{document}

\newcommand{\ourframework}{MirrorNet\xspace}
\newcommand{\backbone}{BackboneNet\xspace}
\newcommand{\hybrid}{MirrorNet\xspace}

\title{\ourframework: A TEE-Friendly Framework for Secure On-device DNN Inference}

\author{\IEEEauthorblockN{Ziyu Liu, Yukui Luo, Shijin Duan, Tong Zhou, Xiaolin Xu}
\IEEEauthorblockA{{Northeastern University, Boston, MA, USA}}
\IEEEauthorblockA{\{\texttt{liu.ziyu4, luo.yuk, duan.s, zhou.tong1, x.xu}\}@northeastern.edu}}

\maketitle

\begin{abstract}
Deep neural network (DNN) models have become prevalent in edge devices for real-time inference. However, they are vulnerable to model extraction attacks and require protection. Existing defense approaches either fail to fully safeguard model confidentiality or result in significant latency issues.
To overcome these challenges, this paper presents \ourframework, which leverages Trusted Execution Environment (TEE) to enable secure on-device DNN inference. It generates a TEE-friendly implementation for any given DNN model to protect
the model confidentiality, while meeting the stringent computation and storage constraints of TEE. The framework consists of two key components: the backbone model (\textit{\backbone}), which is stored in the normal world but achieves lower inference accuracy, and the Companion Partial Monitor (CPM), a lightweight mirrored branch stored in the secure world, preserving model confidentiality. During inference, the CPM monitors the intermediate results from the \backbone and rectifies the classification output to achieve higher accuracy.
To enhance flexibility, \ourframework incorporates two modules: the CPM Strategy Generator, which generates various protection strategies, and the Performance Emulator, which estimates the performance of each strategy and selects the most optimal one.
Extensive experiments demonstrate the effectiveness of \ourframework in providing security guarantees while maintaining low computation latency, making \ourframework a practical and promising solution for secure on-device DNN inference. For the evaluation, \ourframework can achieve a 18.6\% accuracy gap between authenticated and illegal use, while only introducing 0.99\% hardware overhead.
\end{abstract}


\section{Introduction}
Deep neural networks (DNN) are designed to automatically learn the complex pattern and feature representation of the input data, which has been successfully used for various applications, like facial recognition \cite{kortli2020face}, autonomic driving\cite{kuutti2020survey}, and health care monitoring\cite{alazzam2021novel}. However, given concern about the data privacy, many users of the DNN model prefer not to share their private data with the online server. 
As a result, there is a growing trend in implementing DNN models on edge devices \cite{xu2019first}. By deploying DNN models directly on edge devices, such as smartphones, the model inference can be performed locally to protect user privacy,
since most sensitive data remains in device. Moreover, DNN inference on edge can achieve lower latency compared to cloud computing.

Despite its encouraging performance, deploying high-performance DNN models on edge devices is vulnerable to model extraction \cite{sun2021mind}. Specifically, attackers can extract the model architecture and weights, then transplant it to the unauthorized device without claiming ownership or paying the patent fee. However, training a high-performance DNN model requires a large number of labeled data and substantial computational resources \cite{ying2019bench}. Therefore, these high-performance models are the intellectual property (IP) of the model owners and should be well-protected \cite{pawlicki2022survey}.

\begin{table}[t]
    \centering
    \caption{The comparison with previous works}
    \label{tab:comparison}
    \resizebox{\linewidth}{!}{
    \begin{tabular}{l|c|c|c|c}
    \hline
    \hline
        & Privacy & Latency & Flexibility & Accuracy \\ \hline
        
        DarkneTZ \cite{mo2020darknetz} & \emptycirc & \halfcirc & \fullcirc& \fullcirc \\
        eNNclave \cite{schlogl2020ennclave} & \fullcirc & \halfcirc & \emptycirc & \emptycirc\\
        Confidential DL \cite{vannostrand2019confidential} & \fullcirc & \emptycirc & \fullcirc & \fullcirc\\
        ShadowNet \cite{sun2020shadownet} & \halfcirc & \emptycirc & \fullcirc & \fullcirc\\
        \textbf{\ourframework ~(ours)} & \fullcirc& \fullcirc&\fullcirc & \fullcirc \\
        \hline
    \end{tabular}}
    \begin{tablenotes}
        \item{\emptycirc~ Not covered;~~ \fullcirc~ Covered; ~~ \halfcirc~ Partially covered} 
    \end{tablenotes}
\end{table}

Several methods have been proposed for model protection, such as {watermarking \cite{li2021survey}, non-transferable learning \cite{wang2021non}, and obfuscation \cite{zhou2022obfunas}}. Among the existing methods, we believe that Trusted Execution Environment (TEE)-based solutions are particularly well-suited for safeguarding DNN models on edge devices. TEE is commonly available in modern edge devices, such as ARM processors, and
refers to an area (secure world) inside the main processor of the device that is separated from the system's main operating system (normal world). TEE ensures the confidentiality and integrity of data processing, making it a widely utilized technology for user authentication and key management on edge devices \cite{yandamuri2021communication}. 
However, fully deploying DNN models inside TEE to achieve model protection is impractical, due to their increasing model size and the limited computational resources and storage memory within TEEs. For example, the Raspberry Pi 3B with ARM Cortex-A chip offers a maximum of 16MB memory for the TEE \cite{pi2015raspberry}, which is incompatible with the state-of-the-art DNN models, e.g., over 100MB for ResNet-101 \cite{he2016deep}.

Although several previous works employed TEE to secure the DNN inference by uploading certain layers to TEE \cite{mo2020darknetz} or introducing extra masking and linear transformation in TEE between the execution of layers \cite{sun2020shadownet}, these methods either cannot fully protect the model or incur a large overhead, as summarized in Tab.\ref{tab:comparison} and detailed in Sec. \ref{sec:related}. More importantly, all these existing methods fail to address a significant vulnerability associated with layer-wise dependency in DNN computation.
For example, if the first layer is executed in the secure world and the second in the normal world, the intermediate feature maps might be leaked during the communication, and the attacker could retrain the model with less effort.
i.e., merely isolating a few layers within the secure world of TEE does not provide sufficient protection for maintaining model confidentiality (see results in Sec. \ref{sec:re-training_vulnerability}).

To overcome these design challenges, we propose a framework, namely \ourframework,
which aims to protect the functionality of the input DNN model by generating a TEE-friendly deployment for the given model. Additionally, it addresses the vulnerability of model extraction attacks by limiting attackers to only extracting a poorly performing model.
We take the architecture of the input DNN model as a backbone network, which is deployed in the normal world. Besides, we develop a lightweight \textit{mirror network} based on the backbone network, namely Companion Partial Monitor (CPM), which is connected with the backbone but stored in the secure world. Through model training, we enable the entire model to achieve high accuracy, while intentionally degrading the accuracy of the backbone model in the normal world. For clarity and conciseness, we refer to both our framework and the combined model as \ourframework.
In summary, \ourframework addresses the vulnerability of model extraction by safeguarding the lightweight mirror network within secure world. This lightweight network is designed to meet the computational and storage limitations of the TEE. It serves as a crucial component of the entire model, enabling authorized users to perform high-performance inference in conjunction with the backbone network.

The contributions of this work are summarized as follows:
\begin{itemize}
    \item We identify and validate a neglected vulnerability in the existing DNN model protection methods, see Sec. \ref{sec:re-training_vulnerability}. For the first time, we take the re-training vulnerability into consideration in developing a TEE-friendly framework \ourframework, for secure DNN inference on edge devices. 
    
    \item We leverage the layer-wise and channel-wise dependency of DNN models to generate a mirrored network architecture named Companion Partial Monitor (CPM) for a given victim DNN model. As a lightweight network, CPM can be deployed in the limited secure memory of TEE but determines the performance of the overall inference. \ourframework can be optimized with arbitrary strategies and can even be integrated with other purpose training.

    \item We explore numbers of potential mirrored model architectures and equip the \ourframework framework with two components, namely \textit{CPM Strategy Generator} and \textit{Performance Emulator}, to selectively generate the optimal protection scheme. 
    
    \item By deploying \ourframework on a Raspberry Pi 3 Model B board, we validate its performance with input networks including LeNet-5 \cite{lecun1998gradient} and VGG networks \cite{simonyan2014very}. We also train models on three datasets for inference, including MNIST~\cite{lecun1998mnist}, FashionMNIST~\cite{xiao2017fashion}, and CIFAR-10~\cite{krizhevsky2010convolutional}, to evaluate the security level of our method. Experimental results demonstrate that our proposed \ourframework framework achieves comparable accuracy to the input networks with low hardware overhead, e.g., 0.99\% for VGG-7. On the other hand, illegal model extraction from the normal world can only obtain a compromised model, with an accuracy gap as high as 18.6\%.
\end{itemize}

\section{Background and Related Work}

\subsection{Threat Model}
\begin{figure}
    \centering
    \includegraphics[width=1\linewidth]{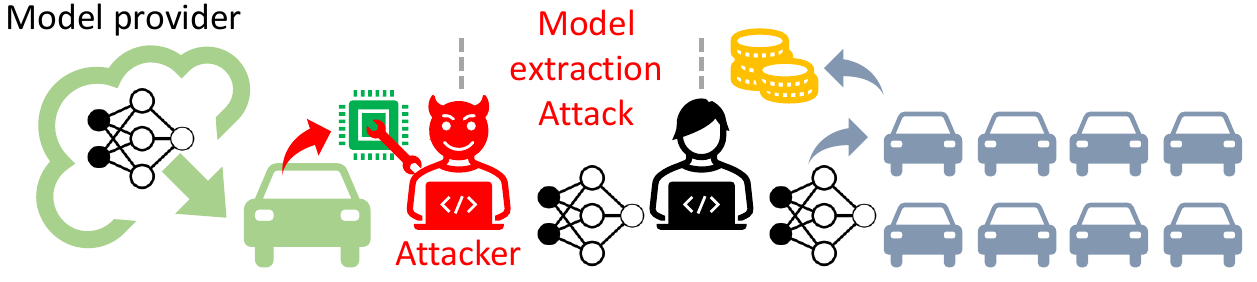}
    \caption{The illustration of our threat model.}
    \label{fig:threat}
\end{figure}

In our thread model, as shown in Fig. \ref{fig:threat}, we focus on real-time scenarios when executing security-sensitive applications, such as a self-driving system or an AI system used for face recognition.
We assume a ``knowledgeable'' attacker whose object is to extract a well-performed model. The attacker can directly acquire sensitive information like the model architecture and weights of a DNN model in the normal world, or correctly infer them from some approaches, such as side-channels \cite{su2021survey}. Besides, we assume the secure world, i.e., the TEE, is isolated and remains as a black box for the attackers, who are not able to infer the computation performed inside the secure world. Yet, the attacker can figure out when the secure world computation is invoked and monitor the data communication between the secure world and the normal world through the shared buffers. Since we focus on the model confidentiality issue during the usage of the TEE-friendly secured model, we assume our \ourframework framework is trained offline and the normal world + secure world implementation is only conducted for the model inference procedure.

\subsection{Trusted Execution Environment (TEE)}

TEE is a secure area within a processor that provides a secure environment for executing code and processing data, guaranteeing confidentiality and integrity inside \cite{jauernig2020trusted}. It helps isolate sensitive operations from the rest of the device and protect it from potential attackers or malware. Taking the ARM TrustZone as an example, it creates two virtual environments, namely \textit{Normal World} and \textit{Secure World}. The normal world is the normal operating environment for most applications. Differently, the secure world has its own isolated memory and peripherals, and provides an environment for running trusted applications with the help of security mechanisms. Although there are some security issues challenging the TEE execution, such as TrustZone \cite{pinto2019demystifying}, it is still the cornerstone of modern secure applications. Moreover, researchers have been dedicated to implementing DNN with the TEE because of its protection of confidentiality and integrity in recent years \cite{vannostrand2019confidential}.

\subsection{Related Works on DNN Protection}
\label{sec:related}
The DNN model execution on CPU has been proven vulnerable to information leakages, such as training data and model parameters \cite{liu2021machine}. In this context, one prominent concern is the model extraction attack, in which the attacker aims to duplicate or ``steal'' the DNN model through different approaches, such as shadow training \cite{tramer2016stealing}, constructing meta-models \cite{oh2019towards}, and exploiting memory and timing side-channels \cite{wang2018stealing}. 

Recent efforts have attempted to utilize TEE for securing DNN inference. However, the limited computation resources and memory in TEE make the implementation rather challenging. DarknetZ\cite{mo2020darknetz} partitions the model and put the last few dense layers inside TEE. However, the convolution layers of the victim model are still running in the normal world in plaintext format, presenting a vulnerability. 
Similarly, another work eNNclave \cite{schlogl2020ennclave} uses a few pre-trained and publicly available convolution layers as the feature extractor in the normal world and a followed user-defined dense layer as the classifier in TEE. However, it suffers from performance loss since the parameters of the feature extractor are fixed during retraining.
Another approach is to protect the whole model with sequential executions. For instance, one work divides the model into partitions and executes each separately inside TEE \cite{vannostrand2019confidential}, which causes an extremely long execution time. 
ShadowNet obfuscates the weights by masking them with linear transformations \cite{sun2020shadownet}. However, the transformation is insecure against ``strong'' attacker who can monitor the memory access pattern and extract the pattern of weights\cite{zhang2022teeslice}.

\section{Challenges of DNN Protection with TEE}

\subsection{Inadequate Confidentiality Protection Against Retraining}\label{sec:re-training_vulnerability}
In previous works like DarknetZ\cite{mo2020darknetz} and eNNclave\cite{schlogl2020ennclave}, although the last few dense layers are protected inside the TEE, the architecture, and parameters of the previous convolution layers are exposed to the attacker in plaintext format. In this situation, an experienced attacker can easily retrain the model with high accuracy and break the protection schemes. 
To be more specific, the attacker can freeze the previous convolution layers which are exactly part of the high-accuracy model, and randomly initializes or customizes a dense layer that is unknown to the attacker. The retraining process will take less effort since the intermediate result after the convolution layers are totally correct. 
A few more computations are needed to obtain the final prediction result.
In order to verify this vulnerability, we trained a LeNet-5 model on the CIAFR-10 dataset for image classification and remove the final dense layer. The result shows that even with random initialization for weights parameter inside the final dense layer, the extracted model was able to be recovered to its original accuracy within just 20 epochs using 1\% of the dataset.

\subsection{Large Hardware Overhead for TEE-assisted Protection}
 
The DNN model typically consists of multiple layers and they are arranged in a sequential manner. During inference, the input data is passed through the model layer by layer and the model forward propagates the result until the final output layer. Previous methods try to split the whole architecture into different parts and run some of them inside TEE. 
However, computation resources are limited inside the TEE and the operation in the TEE runs orders of magnitude slower than in the normal world \cite{tramer2018slalom}. For the example in ShadowNet \cite{sun2020shadownet}, the TEE-assisted model inference time for one image can be ranging from tens to thousands of milliseconds, in terms of different models, and the memory requirement in TEE ranges from several to tens of MegaBytes. This is highly inefficient for resource-limited edge devices.

 \begin{figure*}[!t]
    \centering
    \includegraphics[width=\linewidth]{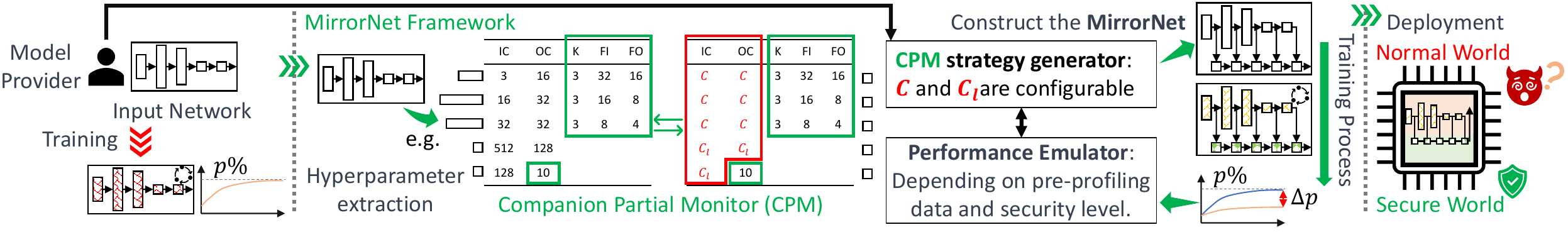}
    \caption{The overview of \ourframework. The model provider selects an input network for the BackboneNet architecture and generates the corresponding CPM. CPM has the same layer type as BackboneNet but with a smaller size. \ourframework is the integration of BackboneNet and CPM. CPM design is generated by a Strategy Generator and evaluated by a Performance Emulator to help the model provider decide if the current design is appropriate for inference performance and hardware deployment. The trained \ourframework is deployed on hardware where BackboneNet is in the normal world and CPM is in the secure world. }

    \label{fig:mirror_process}
\end{figure*}

\subsection{Other Challenges}

The layer-wise feature of a DNN model determines that when leveraging TEE to protect some intermediate layers, the communication between the normal world and the secure world is bidirectional, i.e., the input to the layer in the TEE will return the output to the normal world for further computation. Thus, the intermediate result will be exposed to the attacker who can monitor everything in the normal world, favoring an attacker to infer the function inside the TEE. Accordingly, ShadowNet \cite{sun2020shadownet} adds a mask and linear transformation before sending the intermediate result back to the normal world, which however, increases the computation complexity and data transmission overhead. More severely, the lightweight encryption and masking are prone to be broken \cite{lapid2019cache}.

Regarding the limitations of previous works, a good framework should comprehensively protect the model confidentiality so that the attacker can not extract and transplant the model for unauthorized usage. At the same time, the model protection scheme in the network should not sacrifice the model performance for the legitimate user. The latency overhead should be low for a real-time system and provide a good user experience. Furthermore, the network needs to be generalization which means that it can provide protection for different kinds of model architecture. In this work, we overcome these challenges by proposing \ourframework framework that can adequately address the aforementioned limitations.

\section{Proposed Method: \ourframework}

This section presents \ourframework, which  
transforms an input DNN model (e.g., in Fig. \ref{fig:mirror_process}) into its TEE-friendly counterpart. 
Specifically, \ourframework can achieve a comparable or even better performance, while protecting
model's confidentiality.

\subsection{Preliminaries of DNN Model Architecture}
A representative DNN model is 
composed of multiple interconnected layers, in which the convolutional layer and dense layer are the two most common layers. Taking the 2D convolution layer as an example, 
it is defined by input channels ($IC$), output channels ($OC$), kernel size ($K$), input feature size ($FI$), and output feature size ($FO$). For ease of clarification, we assume the feature maps are square-shaped, i.e., the input/output feature map is $(FI\times FI)$ / $(FO\times FO)$, while other shapes such as rectangular are free to be extended to. Specifically, we define a Conv2D layer as [$IC, OC, K, FI, FO$]. Similarly, the dense layer is defined as [$IC, OC$], indicating its input and output feature dimensions.

\subsection{\ourframework: Overview}

We demonstrate the entire workflow of our \ourframework framework on an edge device in Fig.~\ref{fig:mirror_process}. First of all, the user (e.g., model provider) selects or possesses an input network that needs to be protected, so 
\ourframework generates a BackboneNet with the same architecture as the input network. 
Then, \ourframework builds a mirrored model architecture called Companion Partial Monitor (CPM).
Here, the ``companion partial'' denotes that each layer inside the CPM has the same layer type
as the corresponding layer in the \backbone, but with fewer parameters. For example, for 
a Conv2D layer in  \backbone with size [$3, 64, 5, 32, 28$], its CPM counterpart can be a Conv2D layer with size [$1,1,5,32,28$].
In other words, the CPM
is a scaled-down version (especially channel-wise) of the \backbone,
and its lightweight property well fits the memory constraints inside the TEE. 
We denote the integration of \backbone and CPM as \hybrid, as illustrated in Fig. \ref{fig:mirror_process}.
Subsequently, \ourframework will be trained from scratch. {Rather than initializing the \backbone with pre-trained parameters (if publicly any), randomly initializing the \backbone is supposed to emphasize the importance of CPM more during the \hybrid training.}

With a well-trained \hybrid, the model owner will decide if the current CPM design satisfies design requirements, from both performance (e.g.,
inference latency, model size, etc.) and security (i.e., accuracy gap between \backbone and \hybrid, $\Delta p$ as illustrated in Fig. \ref{fig:mirror_process}). Since there are numerous possible CPM configurations for a BackboneNet, it is less feasible for the user to train them all and find the optimal one. To mitigate this concern, we develop two components for the framework, namely ``CPM Strategy Generator'' and ``Performance Emulator''. As indicated by their names, these two components can estimate the security of a \hybrid and its hardware overhead without practical deployment.

\begin{figure}[t]
    \centering

    \includegraphics[width=0.8\linewidth]{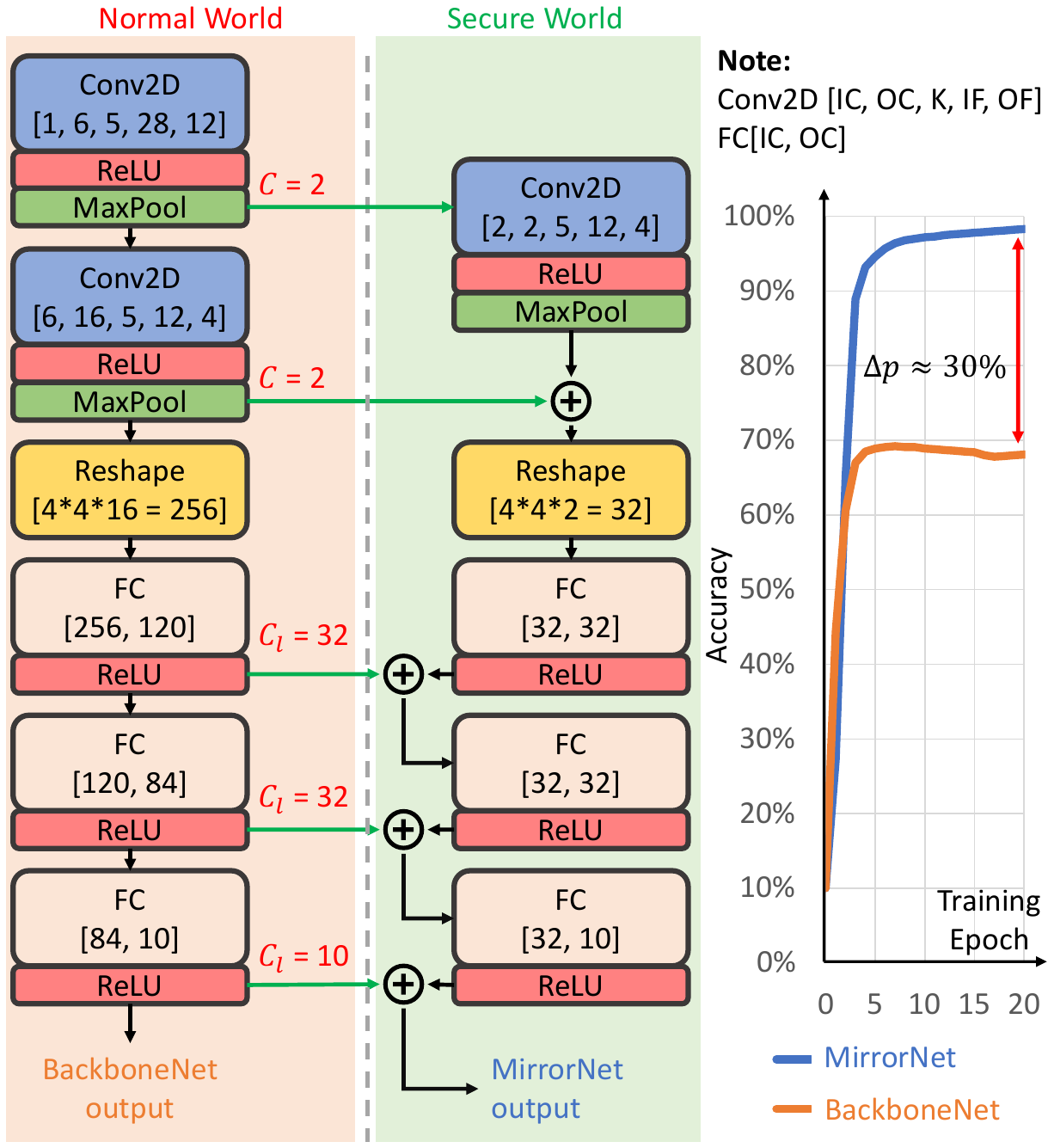}
    \caption{\hybrid example for LeNet-5.}
    \label{fig:lenet5eg}
\end{figure}

\subsection{Construction of \hybrid}
{\ourframework combines \backbone and CPM to execute inference,  
in which we take \backbone as the mainstay and CPM as the auxiliary, and design a \textbf{\textit{feedforward}} route. Each layer's output of \backbone will go into the next layers of both \backbone and CPM, while the layer output of CPM will only be fed into the next CPM layer, i.e., there is no feedback from CPM to \backbone. Taking the first layer as an example, the input enters \backbone, after which part of the output is sent to the first layer in CPM as the input. Note that there is no layer in CPM corresponding to the first layer of \backbone, but the first layer in CPM aligns with the second layer in \backbone. In the separate use of BackboneNet, its output, namely \textit{logits}, will be passed through the \texttt{argmax} function to derive the predicted label for the current input query. In \hybrid, the logits of \backbone are also transferred to TEE and combined with the logits of CPM; and the combination is passed through the \texttt{argmax} function, so that only the predicted label from CPM is transferred back to the normal world indicating the final result.}

To clearly illustrate the workflow, we depict one \hybrid architecture using LeNet-5 \cite{lecun2015lenet} as the \backbone in Fig.~\ref{fig:lenet5eg}.
{The model architecture on the left part stands for the \backbone,  
and the right part is the CPM. 
The layer inside the \hybrid is represented in rectangles with the name (upper part) and the parameter (lower part). ReLU and MaxPooling are represented in a strip shape.} 
The communication between the normal world and the secure world is unidirectional, and the $C$ in between stands for the number of channels in the intermediate result that is transmitted to the secure world. The line graph on the right-hand side of Fig.~\ref{fig:lenet5eg} shows the comparison of inference accuracy between running \backbone only and \hybrid as a whole. 
The result shows that the accuracy gap can be up to 30\%, which is a significant drop when attackers only acquire the \backbone.

\subsection{\ourframework: CPM Strategy Generator}\label{sec:strategy_generator}
For a specific \backbone, there exist many possible CPM configurations. To generate the configurations satisfying the constraint TEE resource, our \ourframework framework embeds a CPM Strategy Generator, as shown in Fig. \ref{fig:mirror_process}. 
It ensures the layer alignment between \backbone and CPM, i.e., the kernel size ($K$), input feature size ($FI$), and output feature size ($FO$ that is related to padding and stride configuration) of CPM convolution layers must stay the same as BackboneNet. For other parameters like input channel ($IC$) and output channel ($OC$), the generator selects a small number to achieve lightweight hardware overhead, denoted as \textcolor{black}{$C$}, which is empirically from 1 to 4 (see Sec. \ref{subsec:layer_latency_profiling}). 
For the dense layer, we determine a small dimension denoted as \textcolor{black}{$C_l$}. Note that $C_l$ is not manually selected, but determined by the $OC$ and $FO$ of the last convolution layer:
\begin{equation}\label{eq:C_setup}
\small
    C_l = C \times FO^2
\end{equation}
{The $C_l$ should not be smaller than the number of classes, i.e., 10 in Fig. \ref{fig:mirror_process}, to ensure well-behaved dense layers. 
While in practice, this condition is typically met.} 

\subsection{\ourframework: Performance Emulator}\label{sec: performance_emulator}
In addition to security, performance overhead is another critical factor in developing TEE-based DNN solutions. Since the performance overhead (i.e., latency) of \hybrid depends on its deployment, similarly, it is infeasible to measure the practical overheads for each possible CPM strategy. Therefore, we build a Performance Emulator to help with the latency estimation. Taking the Cortex-A53 processor used in our experiments as an example, we profile various workloads (e.g., convolutional layers, dense layers, etc.) for the possible configurations of CPM. We collect the hardware latency of these workloads (see Sec. \ref{subsec:layer_latency_profiling}) and run a regression analysis to derive the Performance Emulator. Consequently, with an arbitrary input network (Fig. \ref{fig:mirror_process}), the Performance Emulator can estimate the hardware latency for a specific platform, without actually deploying the entire \hybrid. This will extremely reduce the design efforts when deciding an appropriate \hybrid candidate for a targeted edge device.

By applying the latency profiling results, the CPM 
Performance Emulator can predict a number of \hybrid candidates satisfying the design requirements, 
such as latency overhead. Then, \ourframework can evaluates the accuracy difference $p\%$ between the \backbone and the \hybrid and select the best option according to the privacy requirement.

\subsection{Superiority of \hybrid Performance and Privacy}
\subsubsection{Inference performance}
{As a theoretical analysis, we indicate that the accuracy of the \backbone serves as a lower bound for the accuracy of its corresponding \hybrid. We provide a concise and straightforward proof by setting all the weights in CPM to $0$, so the output of \hybrid is exactly equal to the output of its \backbone.  
Actually, since CPM has a similar structure to the \backbone, the combination of them (i.e., the \hybrid) can be treated as a specially-shaped ensemble model \cite{sagi2018ensemble} of the \backbone. Hence, the \hybrid performance is even potentially improved over its \backbone part. 
However, since the \backbone and CPM are jointly trained, the performance of \backbone alone will be very poor, as shown in the evaluation in Sec.\ref{sec:experiment_validation}.}

\subsubsection{Privacy}

{
An attacker has no information from the secure world since TEE is physically isolated and well-protected for secure computation. Besides, our feedforward design for the \hybrid guarantees that TEE never transmits data to the normal world, except for the final prediction result. Because the CPM computation is between every layer and those protected layers are performed in a cascaded manner in TEE, the attacker can neither infer the intermediate results from the predicted label nor freeze specific layers to retrain the partial model \cite{kolcun2020case}.
Therefore, s/he can only retrain the extracted \backbone from scratch. 
We regard one model extraction attack as failed if the attacker still needs to train the model from scratch.}

\begin{figure*}[t]
    \centering
    \includegraphics[clip, trim=1cm 6.5cm 1cm 6.5cm, width=.9\linewidth]{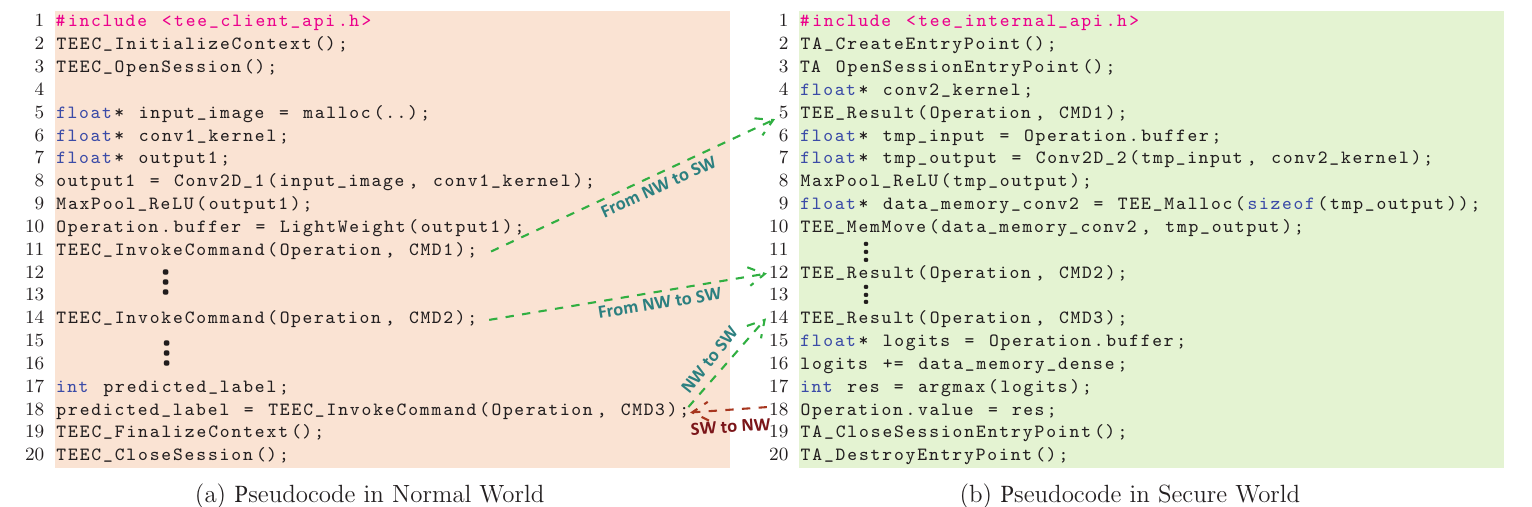}
    \caption{C-based pseudocode for \ourframework implementation, with data transmission between normal world and secure world.}
    \label{fig:Ccode}
\end{figure*}

\subsubsection{Lightweight hardware overhead}
{One important characteristic of \ourframework is that the CPM inside TEE is only a mirrored version of the \backbone with a small size.
Therefore, the computation of CPM will not induce such a long execution time as the \backbone in the normal world, mitigating the limited resource issue in TEE for NN inference. 
The second characteristic is that the computation result of CPM will not be transferred back to the normal world like previous works \cite{sun2020shadownet} but stored in the secure memory.  
Thus, CPM only needs to send a return signal to the normal world, indicating the forward propagation to continue for the followed \backbone layer\footnote{This done signal is an indicator for the completeness of one layer in CPM, which favors the normal world computing schedule if \hybrid is deployed on a multi-thread CPU or a parallel-computing platform. 
}. As the latency between switching worlds can be heavy, our feedforward strategy presents an efficient way, i.e., there is no data communication from secure world to normal world, which greatly reduces the communication overhead during inference; thanks to the already-ensured privacy, no encryption/decryption is needed anymore which further lowers the execution time and the complexity.}

\subsection{\hybrid: Hardware Implementation}

\begin{figure}[t]
    \centering
    \includegraphics[width=.9\linewidth]{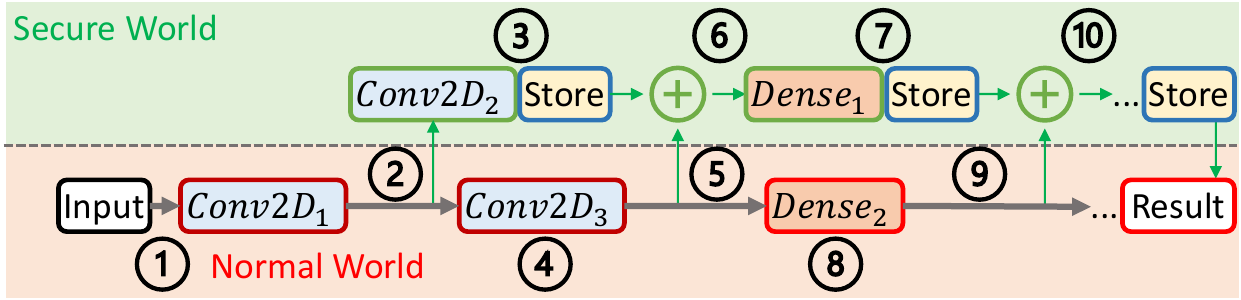}
    \caption{Hardware implementation of \hybrid. 
    ReLU and MaxPooling are omitted for clarity. 
    }
    \label{fig:hw_imp}
\end{figure}

This section presents the hardware implementation of \hybrid. 
We illustrate this procedure by taking a simple convolutional neural network (CNN) as an example, as shown in Fig. \ref{fig:hw_imp}. This CNN is a \backbone that only has two Conv2D layers and one dense layer, and the CPM is composed of one down-scaled Conv2D and one dense layer.
The execution order is interpreted as running on a single-thread CPU equipped with TEE. For other platforms, such as multi-thread CPU or TEE-embedded GPU\footnote{One example for its commercial product is NVIDIA H100 GPU.} \cite{volos2018graviton}, the secure world and normal world can execute their components in parallel and for batched queries as long as satisfying the dependency.

\begin{table*}[!]
\centering
\caption{Apply LeNet-5 as the input network in \ourframework with varying $C$ settings (Fig.\ref{fig:lenet5eg} and Eq. \ref{eq:C_setup}) and different datasets. The corresponding \hybrid is trained to obtain $\Delta p$ and emulate the performance. The CPM latency emulation is on MNIST.}
\begin{tabular}{c|ccc|ccc|ccc|c}
\hline
\multirow{3}{*}{\begin{tabular}[c]{@{}c@{}}Channel\\ number\\ ($C$)\end{tabular}} & \multicolumn{3}{c|}{MNIST} & \multicolumn{3}{c|}{FashionMNIST} & \multicolumn{3}{c|}{CIFAR-10} & \multirow{3}{*}{\begin{tabular}[c]{@{}c@{}}CPM\\ performance\\ Emulator\\ (ms)\end{tabular}} \\ \cline{2-10}
 & \multicolumn{2}{c|}{Input Network Test Acc.(\%)} & \multicolumn{1}{c|}{98.7} & \multicolumn{2}{c|}{Input Network Test Acc.(\%)} & \multicolumn{1}{c|}{89.2} & \multicolumn{2}{c|}{Input Network Test Acc.(\%)} & 62.8 &  \\ \cline{2-10}
 & \begin{tabular}[c]{@{}c@{}}BackboneNet\\ Test Acc.(\%)\end{tabular} & \begin{tabular}[c]{@{}c@{}}\hybrid\\ Test Acc.(\%)\end{tabular} & \begin{tabular}[c]{@{}c@{}}$\Delta p$\\ (\%)\end{tabular} & \begin{tabular}[c]{@{}c@{}}BackboneNet\\ Test Acc.(\%)\end{tabular} & \begin{tabular}[c]{@{}c@{}}\hybrid\\ Test Acc.(\%)\end{tabular} & \begin{tabular}[c]{@{}c@{}}$\Delta p$\\ (\%)\end{tabular} & \begin{tabular}[c]{@{}c@{}}BackboneNet\\ Test Acc.(\%)\end{tabular} & \begin{tabular}[c]{@{}c@{}}\hybrid\\ Test Acc.(\%)\end{tabular} & \begin{tabular}[c]{@{}c@{}}$\Delta p$ \\ (\%)\end{tabular} &  \\ \hline
1 & 62.2 & 98.4 & 36.2 & 57.1 & 88.8 & 31.7 & 39.6 & 62.7 & 23.1 & 3.05 \\
2 & 54.4 & 98.3 & 43.9 & 55.1 & 88.2 & 33.1 & 37 & 62.5 & 25.5 & 3.2 \\
3 & 53.5 & 98.6 & 45.1 & 53.3 & 88.1 & 34.8 & 36.7 & 62.8 & 26.1 & 3.36 \\
4 & 40.5 & 98.2 & 57.7 & 51.5 & 89.2 & 37.7 & 35.8 & 63.1 & 27.3 & 3.42 \\ \hline
\end{tabular}
\label{tab:performance}
\end{table*}

{The execution flow of \hybrid can be summarized as follows: \circled{1} The input query is provided to the normal world that runs the first convolution layer $Conv2D_1$.
\circled{2} The normal world sends certain output channels of the feature map from $Conv2D_1$ to the secure world, which is a small part of the output. \circled{3} The secure world receives the down-scaled feature map and executes the designed small convolution $Conv2D_2$ for feature extraction. The results are then stored in the secure memory in the secure world, i.e., the TEE\footnote{Note that for parallel computing or multi-thread cases, a \textit{done} signal should be returned to the normal world to indicate the end of secure world execution.}. Steps \circled{4} and \circled{5} execute the same procedure as the previous one that runs $Conv2D_3$ and transfers partial results to the secure world. In step \circled{6}, the received feature map from $Conv2D_3$ is added with the stored result of $Conv2D_2$ in TEE, as the amended input feature map for the followed $Dense_1$ layer. To ensure the results from $Conv2D_2$ and $Conv2D_3$ are added, \hybrid requires the output channel of $Conv2D_2$ equal to the number of transferred channels. Further, $Conv2D_2$ and $Conv2D_3$ should have the same configurations on convolution, such that they can produce a feature map with the same size. We omit the description of steps \circled{7}, \circled{8}, and \circled{9}, since they have similar operations except for the dense layer computation. In step \smallcircled{10}, the result from $Dense_1$ and $Dense_2$ are summed up, followed by the argmax function to determine the inferred label for the input query, which is then sent back to the normal world. }

{In addition to the coarse-grained scheduling for \hybrid execution, we further take the open-source OP-TEE \cite{optee} as an example, to illustrate the fine-grained operations with a C program pseudocode, as shown in Fig. \ref{fig:Ccode}.
For the \backbone execution, the memory allocation and variable initialization are all done in the normal world. The followed computations (i.e., $Conv2D_1$) are executed sequentially while the output is partially copied to an \texttt{Operation} structure, which is uploaded to secure world together with the operation command \texttt{CMD1}. Other layers in the normal world follow the same schedule, while the last command will return a result of the entire \hybrid, as the \texttt{predicted\_label}. In the secure world, corresponding commands upon receiving operations from the normal world will be executed (i.e., $Conv2D_2$). 
The memory allocation inside TEE, as shown in Line 9 and 10, utilizes \texttt{TEE\_Malloc} or \texttt{TEE\_MemMove} function. Thus, the data are stored in secure world without leaking any information to the normal world. In each layer, the input buffer copies data from \texttt{Operation.buffer} that is transferred from the normal world and cooperates with the stored data from the output of
previous layer. Note that all the operations, e.g., \texttt{Conv2D}, are defined separately in the normal world and secure world, ensuring the
isolation between them except for necessary data transmission.}

\section{Experimental Validation}\label{sec:experiment_validation}

\subsection{Experimental Setup}\label{sec:experiment_setup}

\subsubsection{Architectures and Datasets}
Since this work focuses on developing a TEE-friendly framework for secure DNN inference, 
our targeted scenarios are lightweight applications rather than complicated tasks like ILSVRC image classification \cite{russakovsky2015imagenet}. Moreover, our proposed framework
offers fine-grained model protection at the layer-wise level. This makes it versatile to be generally applied to any DNN model 
To demonstrate its practicality, in our experiment, we use popular benchmarks on resource-limited edge devices, as the proof-of-concept. Specifically, we start from an easy task, MNIST \cite{mnist}, and then extend to more practical and complicated tasks like FashionMNIST \cite{FashionMNIST} and CIFAR-10 \cite{cifar}. 
For the architecture, we evaluate \ourframework based on LeNet-5 \cite{lecun1998gradient} and VGG \cite{simonyan2014very}.

\begin{figure*}[!]
    \centering
    \includegraphics[width=1\linewidth]{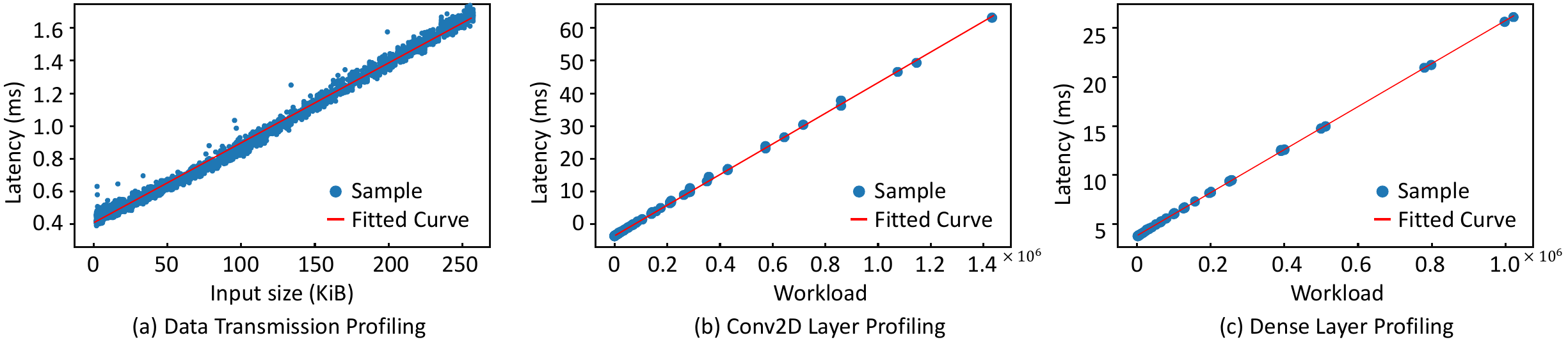}
    \caption{The execution time of different components in \ourframework.}
    \label{fig:latency}
\end{figure*}

\subsubsection{Implementation}
{We implement \ourframework in C language on a Raspberry Pi 3 Model-B, which has a BCM2837 64bit ARM Cortex-A53 Quad Core Processor and 1GB RAM.} 
Without loss of generality, we leverage the Open Portable TEE (OP-TEE) \cite{optee}, an open-source TEE framework supported by different boards equipped with Arm TrustZone \cite{TrustZone}. 
OP-TEE provides a secure area in a processor, Internal Core API for Trusted Applications, and the TEE Client API to communicate with TEE.
Note that although we use the ARM TrustZone as an example, our proposed \ourframework framework can be generally applied to other TEEs, e.g., Intel SGX \cite{costan2016intel}, AMD SEV \cite{sev2020strengthening}, Sanctum \cite{costan2016sanctum}, and Sanctuary \cite{brasser2019sanctuary}.

\subsection{Inference Performance}\label{sec:inference_performance}
We first investigate the inference accuracy of \hybrid. 
Note that we do not specify training strategies of \hybrid, since it is capable of employing arbitrary strategies developed in the deep learning regime.
While we preserve the flexibility of \hybrid training, we evaluate \ourframework with the basic strategy, i.e., {Adam \cite{Adam} as the optimizer and Cross-Entropy \cite{crossentropy} as the loss function}. We demonstrate our inference performance results of \hybrid and its \backbone in Tab.\ref{tab:performance}, using LeNet-5.

From Tab.\ref{tab:performance} we can observe that for all datasets, the inference accuracy of \hybrid is very close to the input network, which aligns with our expectations. Even with only one channel (i.e., $C=1$), the \hybrid can achieve an accuracy as good as the input network; as the channel number increases, the \hybrid accuracy gets improve as well. For instance, the \hybrid of $C=4$ on CIFAR-10 even achieves better accuracy than the input network. On the other hand, with the help of CPM, the directly extracted \backbone performs much worse than the \hybrid. The accuracy gap is large enough to make the \backbone useless if the attacker wants to transplant it to another unauthorized device for direct use. As the channel number increases, the accuracy gap between \backbone and \hybrid also increases.

{Further, we provide a detailed case study for the evaluation of the CPM Strategy Generator (Sec. \ref{sec:strategy_generator}). We apply various input networks of different VGG configurations and also vary the channel number $C$ to investigate the accuracy changes, in Tab.\ref{tab:CPM_evaluation}.  
It is evident that the channel number increment can 
increase the accuracy gap $\Delta p$ between \hybrid and \backbone. More importantly, as the complexity of the input network gets larger, although the \hybrid performance becomes better, the accuracy gap actually decreases. From VGG-6 to VGG-16, the \hybrid accuracy increases from 86.1\% to 94.7\%, but the accuracy gap decreases from 22.8\% to 8.6\%, when $C=4$. This is a trade-off when the model provider considers the CPM design and the input network selection. While indeed a more complex input network can lead to a better performance \hybrid, it could harm security by reducing the accuracy gap. If the model owner intends to increase the channel number for better security, the hardware overhead should be considered.}

\begin{table}[t]
\centering
\caption{Example \ourframework strategies for various input networks on the CIFAR-10 dataset.}
\resizebox{\linewidth}{!}{
\begin{tabular}{cc||cccc}
\hline
\begin{tabular}[c]{@{}c@{}}Input\\ Network\end{tabular} & \begin{tabular}[c]{@{}c@{}} Test Acc. \\ (\%)\end{tabular} & \begin{tabular}[c]{@{}c@{}}Channel\\ \# ($C$)\end{tabular} & \begin{tabular}[c]{@{}c@{}}\backbone\\ Test Acc.  (\%)\end{tabular} & \begin{tabular}[c]{@{}c@{}}\hybrid\\ Test Acc. (\%)\end{tabular} & \begin{tabular}[c]{@{}c@{}}$\Delta p$\\ (\%)\end{tabular} \\ \hline
\multirow{4}{*}{VGG-6} & \multirow{4}{*}{85} & 1 & 84.2 & 85.4 & 1.2 \\
 &  & 2 & 81.4 & 86.5 & 5.1 \\
 &  & 3 & 69.6 & 85.8 & 16.2 \\
 &  & 4 & 63.3 & 86.1 & 22.8 \\ \hline
\multirow{4}{*}{VGG-7} & \multirow{4}{*}{91.3} & 1 & 91.0 & 91.2 & 0.2 \\
 &  & 2 & 87.2 & 91.5 & 4.3 \\
 &  & 3 & 80.8 & 91.2 & 10.4 \\
 &  & 4 & 73.3 & 91.9 & 18.6 \\ \hline
\multirow{4}{*}{VGG-11} & \multirow{4}{*}{92.1} & 1 & 91.7 & 92.0 & 0.3 \\
 &  & 2 & 91.6 & 92.2 & 0.6 \\
 &  & 3 & 87.5 & 92.1 & 4.6 \\
 &  & 4 & 79.7 & 92.3 & 12.6 \\ \hline
\multirow{4}{*}{VGG-16} & \multirow{4}{*}{94.6} & 1 & 94.3 & 94.5 & 0.2 \\
 &  & 2 & 93.6 & 94.4 & 0.8 \\
 &  & 3 & 92.0 & 94.5 & 2.5 \\
 &  & 4 & 86.1 & 94.7 & 8.6 \\ \hline
\end{tabular}}
\label{tab:CPM_evaluation}
\end{table}

\subsection{Latency Profiling for Performance Emulator}\label{subsec:layer_latency_profiling}
To use Performance Emulator (Sec. \ref{sec: performance_emulator}), we
separately measure the latency of different operations, such as 
the context switch between the secure and normal worlds with data communication, and the computation in the secure world.

\subsubsection{Context switch}
Due to the dependency between different model layers, the interaction (i.e., data exchange or communication) between the normal world and the secure world is inevitable, which is time-consuming while the latency
also depends on the transferred data buffer size. We first measure the relationship between the input size and the context switching latency, for which we send input data of different sizes to the operation buffer, invoking the CPM and making it return immediately. We record the execution time as shown in Fig. \ref{fig:latency}(a). With the increasing size of data, the latency of context switching also increases accordingly. This result highlights another advantage of \ourframework, which does not require CPM in the secure world to send data back to the normal world after computation, leading to 
latency reduction. 

Following the results in Fig. \ref{fig:latency}, we figure out the linear relationship between the context switching latency and the data size, in Eq. \ref{eq:latency_linear}. Note that the latency is in microseconds ($\times10^{-6}$ sec) and the input size is in kilobytes.
\begin{equation}\label{eq:latency_linear}\small
    \text{Context Switching Latency} \approx 4.888 \times Input Size + 345.9
\end{equation}

\subsubsection{Workload latency inside TEE} {As the computing resource of TEE is limited and fixed, the execution workload (\# of required operations) of certain components dominates the latency of each layer. 
Specifically, we investigate two critical layers in TEE, the Conv2D layer and the Dense layer. Other layers like ReLU consume much shorter time than these two modules, thus we focus on the most time-consuming parts during TEE computations.}
{The workload of convolution stands for the number of operations required for a Conv2D layer, including matrix multiplication and addition that process the input with kernels and generate the output. It is measured in terms of floating-point operations. The workload depends on the specific layer architecture. We provide a general formula to calculate the workload for a typical convolution layer. Recall the notation of a Conv2D layer as [$IC, OC, K, FI, FO$], the workload can be calculated with Eq. \ref{eq:conv_workload}.
\begin{equation}\label{eq:conv_workload}\small
    \text{Workload} \approx K^2 \cdot IC \cdot OC \cdot FO^2 
\end{equation}
We define one multiplication and one addition as a workload unit for Conv2D. Similarly, we define the dense layer workload as the multiplication-addition counts, i.e., Workload $\approx IC\times OC$ for a dense layer.} With the layer-wise latency profiling, we measure the latency (microseconds) as
\begin{equation}\label{eq:convolution_latency}\small
    \text{Conv2D Latency} \approx 0.03938 \times Workload + 504.3
\end{equation}
\begin{equation}\label{eq:dense_latency}\small
    \text{Dense Latency} \approx 0.02666 \times Workload + 465.6 
\end{equation}

Interestingly, the experimental results in Fig. \ref{fig:latency} indicate the linear relationship between different workloads/operations and their latency. Therefore, we can simply apply the formulation in Eq. \ref{eq:latency_linear}, \ref{eq:convolution_latency}, and \ref{eq:dense_latency} to 
configure the Performance Emulator and estimate the performance of a specific CPM strategy (Sec. \ref{sec:strategy_generator}), as well as make 
a trade-off between latency and privacy.

\subsection{End-to-end Performance Evaluation}

We evaluate the performance of different input models protected by \ourframework by measuring their practical end-to-end execution time, the results as shown in Tab. \ref{tab:over_all_performance}. We first measure the execution time of the input network, i.e., without any protection. Then we measure the total execution time of the model inference using \ourframework and 
calculate the overhead. After that, we predict the execution time of the CPM using the Performance Emulator. From these experiments, we observe substantial memory movement and allocation associated with the convolution computations, which are challenging to quantify. To facilitate the performance emulation, we empirically applied a scaling factor 1.6 (following experimental results in VGG-7), to refine the latency estimation. We apply this scaling factor on other models for validation. The experimental results demonstrate that our Performance Emulator achieves high prediction accuracy. Note that the emulation error in LeNet-5 architecture appears larger, due to its original lower latency overhead, i.e., even a minor estimation error (e.g., 2.72 $ms$)
can magnify in percentage terms. In practice, our emulator just needs to provide a coarse-grained prediction for the latency to determine a few optimal CPM strategies.

\begin{table}[!]
\centering
\caption{\ourframework Performance Emulator prediction vs. measurement for different architectures, where $C$ = 4.}

\resizebox{\linewidth}{!}{
\begin{tabular}{cc||cc|cc}
\hline
\begin{tabular}[c]{@{}c@{}}Inpute\\ Network\end{tabular} & \begin{tabular}[c]{@{}c@{}} Measured \\ latency (ms)\end{tabular} & \begin{tabular}[c]{@{}c@{}}\hybrid \\ Measured\\ latency (ms)\end{tabular} & \begin{tabular}[c]{@{}c@{}}Performance\\ Overhead\\ (\%)\end{tabular} & \begin{tabular}[c]{@{}c@{}}CPM \\ Emulator\\Predicted (ms)\end{tabular} & \begin{tabular}[c]{@{}c@{}}Emulation \\ Error\\ (\%)\end{tabular} \\ \hline
LeNet-5 & 8.26 & 10.98 & 32.93 & 3.42&  25.76\\
VGG-7 & 716.559 & 723.719 & 0.99 & 6.960 & 2.80  \\
VGG-11 & 1246.138 & 1255.06  & 0.72 & 8.495 &  4.78\\
VGG-16 & 4530.198 & 4552.154 & 0.48 & 22.061 & 0.48  \\
\hline
\end{tabular}\label{tab:over_all_performance}}
\end{table}

\section{Conclusion}
This paper presents \ourframework for secure DNN inference on edge devices. As a Trusted Execution Environment (TEE) friendly framework, \ourframework 
constructively converts an input DNN model into two parts: a \backbone and a companion partial monitor (CPM). Specifically, the \backbone part is stored in the normal world and only retains poor performance, which is rectified to high performance by the CPM. To make \ourframework more flexible for any input, we propose two components, CPM Strategy Generator and Performance Emulator. Experimental results on a Raspberry Pi demonstrate the good performance and practical applicability of the proposed framework and its complementary components. 

\section*{Acknowledgement}

This work is supported in part by the U.S. National Science Foundation under Grants OAC-2319962, CNS-2239672, CNS-2153690, CNS-2326597, and CNS-2247892.

\bibliographystyle{IEEEtran}
\bibliography{ref}

\end{document}